\documentclass[conference,hyphens]{IEEEtran}
\usepackage{hyperref}
\hypersetup{ 
    colorlinks=true,
    linkcolor=blue,
    filecolor=blue,      
    urlcolor=blue,
}
\usepackage{array, makecell}
\usepackage[inline]{enumitem}
\usepackage{cite}
\usepackage{caption}
\usepackage{amsmath,amssymb,amsfonts}
\usepackage{float}
\usepackage[caption=false, font=footnotesize]{subfig}
\usepackage[font={small}]{caption}
\usepackage{algorithm, algorithmic}
\usepackage{booktabs}
\usepackage{tikz}
\usepackage{pgfplots,filecontents}
\usepackage{cprotect}
\usepackage{mathtools}
\usepackage{xpatch}
\usepackage{xcolor}
\usepackage{realboxes}
\usepackage{verbatim}
\usepackage{bm}
\usepackage{multirow}
\usepackage{listings}
\usepackage{soul}
\usepackage{orcidlink}
\makeatletter
\xpretocmd\lstinline{\bgroup\appto\lst@DeInit{\egroup}}{}{}
\makeatother
\usepackage{graphicx}
\usepackage[binary-units]{siunitx}
\DeclareSIUnit\event{event}
\usepackage{textcomp}
\usepackage{xcolor}
\usepackage{paralist}
\usepackage{soul}
\usepackage[caption=false]{subfig}
\def\BibTeX{{\rm B\kern-.05em{\sc i\kern-.025em b}\kern-.08em
    T\kern-.1667em\lower.7ex\hbox{E}\kern-.125emX}}
\newcommand*\circled[1]{%
  \tikz[baseline=(C.base)]\node[draw,circle,inner sep=0.6pt,line width=0.4mm,](C) {#1};}

\newcommand{\upp}{\vspace*{-0.5em}}
\begin{document}
\title{
   eBPF-Based Instrumentation for Generalisable Diagnosis of Performance Degradation \upp
}

\author{\IEEEauthorblockN{Diogo Landau\orcidlink{0009-0000-2625-8884
}\IEEEauthorrefmark{1},
Jorge Barbosa\orcidlink{0000-0003-4135-2347}\IEEEauthorrefmark{2}, Nishant Saurabh\orcidlink{0000-0002-1926-4693}\IEEEauthorrefmark{1}} 
\IEEEauthorblockA{\IEEEauthorrefmark{1}\textit{Department of Information and Computing Sciences, Utrecht University, NL}}
\IEEEauthorrefmark{2}{\textit{LIACC, Faculdade de Engenharia da Universidade do Porto, Portugal}}\\
d.hewittmouracarrapatolandau@uu.nl\IEEEauthorrefmark{1},
jbarbosa@fe.up.pt\IEEEauthorrefmark{2},
n.saurabh@uu.nl\IEEEauthorrefmark{1}
\upp\upp\upp
}

\maketitle

\begin{abstract}
    Online Data Intensive applications (e.g. message brokers, ML inference and databases) are core components of the modern internet, providing critical functionalities to connecting services. The load variability and interference they experience are generally the main causes of Quality of Service (QoS) degradation, harming depending applications, and resulting in an impaired end-user experience. Uncovering the cause of QoS degradation requires detailed instrumentation of an application's  activity. Existing generalisable approaches utilise readily available system metrics that encode interference in kernel metrics, but unfortunately, these approaches lack the required detail to pinpoint granular causes of performance degradation (e.g., lock, disk and CPU contention). In contrast, this paper explores the use of fine-grained system-level metrics to facilitate an application-agnostic diagnosis of QoS degradation. To this end, we introduce and implement $16$ \textit{eBPF-based metrics} spanning over six kernel subsystems, which capture statistics over kernel events that often highlight obstacles impeding an application's progress. We demonstrate the use of our \textit{eBPF-based metrics} through extensive experiments containing a representative set of online data-intensive applications. Results show that the implemented metrics can deconstruct performance degradation when applications face variable workload patterns and common resource contention scenarios, while also revealing applications' internal architecture constraints.
\end{abstract}


\maketitle

\section{Introduction}
The modern internet is often powered by a vast network of  interconnected applications (e.g., online data-intensive applications ~\cite{meisner2011power, sriraman2018mutune} such as databases, message brokers, and machine learning solutions), providing data, services, and functionality to connecting processes ~\cite{sriraman2018mutune}. Load variability and interference can degrade these applications' Quality of Service (QoS), negatively affecting dependent applications and end-user experience~~\cite{manghwani2005end,li2020amoeba, arapakis2014impact, bouch2000quality}. Restoring QoS requires diagnosing the cause of degradation to apply suitable mitigation ~\cite{dean2014perfcompass}. This necessitates a solution capable of \textit{observing} an application's \textit{performance} by collecting appropriate metrics to understand the system's state.

Unfortunately, attempts at attaining performance observability across diverse application types faces challenges due to differences in threading models ~\cite{sriraman2018mutune}, inter-process communication ~\cite{kerrisk2010linux}, domain-specific metrics ~\cite{redis_exporter,mysql_exporter,kafka_exporter}, and userspace virtual machines ~\cite{node_usdt, jvm_usdt}. These factors impede \textit{generalisability}, hindering the development of widely applicable solutions.

\begin{figure}[hbt!]
    \centering
    \upp\upp\upp\upp
    \subfloat[\label{fig:database:latency}]{
      \includegraphics[width=0.31\columnwidth]{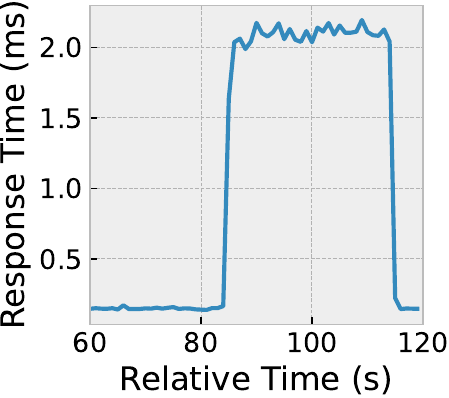} 
    }
    \subfloat[\label{fig:database:lock_wait}]{
      \includegraphics[width=0.31\columnwidth]{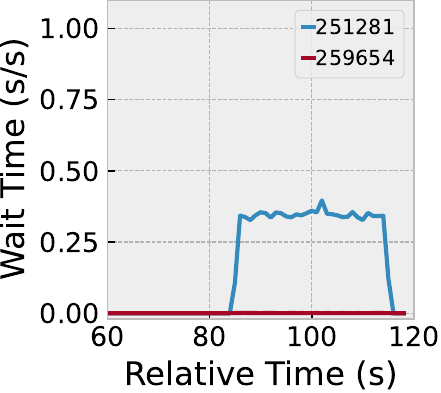} 
    }
    \subfloat[\label{fig:database:lock_wake}]{
      \includegraphics[width=0.30\columnwidth]{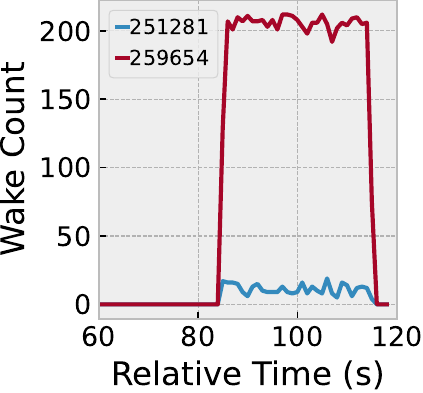} 
    }\upp 
    \caption{\textbf{Illustration of a database experiencing lock contention, leading to degraded performance:} (a) $95^{th}$ percentile response time; (b) Database threads wait time for a lock; (c) Database threads wake activity for a lock.}
    \upp\upp
\end{figure}

Generalisable approaches to estimate an application's state of performance~~\cite{grohmann2019monitorless, cheng2024fedge, yanggratoke2015predicting, ahmed2018automated} and to perform root-cause identification~~\cite{cohen2004correlating, zhang2005ensembles, gan2021sage, gan2019seer}, have focused on leveraging readily-available metrics (e.g. \texttt{proc} filesystem) to analyse the system's state. However, these approaches neglect important \textit{finer-granularity} metrics that can enable more precise diagnosis. For example, consider a relational database that has to serialise read and write operations on a database row. Figures ~\ref{fig:database:latency}, ~\ref{fig:database:lock_wait}, and ~\ref{fig:database:lock_wake} show the database's response time and two threads' lock wait and wake activities. The response time increase clearly correlates with lock activity. Aforementioned works, lacking fine-grained thread-resource interaction monitoring (e.g. \textbf{``thread$\rightarrow$lock''}), can't diagnose issues where the constraint is concealed by a thread's use of specific kernel resources. 

A notable first effort in this direction is presented by ~\cite{seo2022nosql} which creates a generalisable set of metrics to diagnose performance degradation observed in NoSQL applications. Their approach, however, only focuses on NoSQL databases, whereas we aim to extend applicability to a broader range of applications. Moreover, there is a delicate balance between degradation diagnosis and overhead: collecting metrics at too fine a granularity can cause high overhead, while reducing granularity to lower overhead may compromise successful performance degradation diagnosis.

As such, this paper aims to answer the question: ``\textit{Which metrics can be collected to support application-agnostic performance degradation diagnosis, while balancing diagnosability and the performance overhead introduced by granular observability?}''. 

To achieve this, we introduce and implement $16$ \textit{eBPF-based metrics} across six kernel subsystems, including scheduling, virtual file system (VFS), networking, futex, multiplexing IO, and block IO. These metrics enhance observability by tracking a thread's use of specific kernel resources, revealing common degradation scenarios like lock, disk, CPU, and service dependency contention. Furthermore, the granularity of these metrics allows selective thread tracking, narrowing the search space to only those threads that directly or indirectly interact with the entrypoint threads throughout the degradation period.

To validate the implemented metrics, we conducted an extensive series of experiments using a representative set of online data-intensive applications, offering variability in application domains, threading models, inter-process communication (IPC) methods, and user-space virtual machines. Our experiments considers scenarios where applications degrade under a combination of variable workload patterns, using benchmarks and different types of resource contention. 

Our results show that the behaviour leading to the application's degradation was captured by one or more of the implemented metrics in each degradation scenario. However, our method is restricted to Linux systems and has limitations in diagnosing memory contention scenarios. Currently, memory contention is represented as increased CPU runtime (resulting from stalls), rather than variations in expected hardware performance counters (such as $L1$ cache miss, TLB miss). We intend to address this limitation in our future work. 

Nevertheless, in lock contention scenarios, the metrics enabled identifying the threads and futex resources hindering application response time. For disk contention, they revealed the application's inter-thread communication model, linking server response time to disk performance. In CPU-constrained cases, the implemented metrics highlighted impaired threads, identifying runqueue activity as the primary cause of degraded performance. Finally, when degradation originated from an external service dependency, metrics revealed the external service impairing performance, indicating the issue was outside the application under analysis. The complete source code of instrumented metrics and reproducibility artifact is released opensource in an \emph{online} \verb|GitHub| repository~\cite{Prism}.

\section{Related Work}\label{sec:related}
This section discusses state-of-the-art system instrumentation methods employed in KPI estimation and root-cause identification related works.

\paragraph{\textbf{System-level Instrumentation}} 
Application-agnostic performance monitoring methods collect system-level metrics to understand the application's interaction with the kernel, virtualized resources or the underlying hardware. One such method is sampling, which collects system-level metrics at periodic intervals to derive statistics encoding an application's activity~\cite{nagar2007your, gavin1998performance}. Another method is binary instrumentation~\cite{kreutzer2023runtime, seo2022nosql, wang2024diagnosing}, which dynamically inserts probes in the application to collect custom performance statistics. 

\emph{SystemTap}~\cite{prasad2005locating}, \emph{LTTng}~\cite{desnoyers2006lttng} and \emph{Dtrace}~\cite{cantrill2004dynamic} are early solutions which dynamically instrument kernel-level code through dedicated kernel modules. However, their unrestricted kernel memory access frequently triggered kernel panics, rendering them unsuitable for production environments. Nevertheless, \emph{Extended Berkeley Packet Filter (eBPF)}~\cite{ebpf} addressed this issue by implementing and utilizing a kernel virtual machine which restricts \emph{ebpf} program's memory access through specialised helper functions, while interpreting probe instructions. 

Typically, \emph{eBPF} probes are event-driven lightweight programs that are placed by specific system calls into kernel memory, and executed in a protected environment. Such programs are often difficult to develop, primarily due to its complex interactions with kernel data structures, events and its sandboxed execution. However, tools like \emph{BCC}~\cite{bcc}, \emph{bpftrace}~\cite{bpftrace}, \emph{libbpf}~\cite{libbpf} and \emph{libbpf-rs}~\cite{libbpfrs} abstract \emph{eBPF's} low-level details and provide a high-level interface to develop \emph{ebpf} programs. In this paper, we use \emph{libbpf} \cite{bpftrace} to strategically trace particular kernel functions (see details in Section \ref{sec:methodology}).

\paragraph{\textbf{Instrumentation in Application KPI Estimation}} 
A common application of low-level system instrumentation is to estimate high-level key performance indicators (KPIs), while providing a generalised approach to a set of characteristically different applications~\cite{yokelson2024enabling}.
Grohmann et al.~\cite{grohmann2019monitorless} proposed the \emph{Monitorless} framework, which leverages the \emph{Utilization, Saturation and Error (USE)}~\cite{gregg2013thinking} method and samples light-weight platform-level metrics from a host's operating system, using \emph{Performance Co-Pilot (PCP)}~\cite{pcp}. 
However, \emph{Monitorless} relies on a set of coarse-grained system-level metrics (e.\,g., CPU and memory utilization, disk I/O), and 
hence, lacks granularity to capture all application-specific performance issues, not reflected in standard metrics.

Similarly, Cheng et al. developed \emph{FEDGE}~\cite{cheng2024fedge} framework which explicitly focuses on resource contention and collects VM and hardware-level statistics to estimate performance of an application co-located with other services. 
Few other works~\cite{yanggratoke2015predicting, ahmed2018automated} follow a similar approach in the context of video-streaming service's performance estimation. Systems proposed in both these works require monitoring thousands of kernel-level metrics and large numbers of samples. Such performance diagnosis techniques are practically infeasible to employ in real-world systems due to their large computational overheads.

Rezvani et al.~\cite{rezvani2024characterizing} use \emph{eBPF} programs to collect system-level metrics for estimating application's performance state in terms of latency, saturation, and slack. However, their narrow focus on networking and epoll subsystems limits observability of constraints from other subsystems, such as futex and block IO. Similarly, Jha et al \cite{jha2022holistic} leverage \emph{eBPF}-collected metrics to troubleshoot performance issues like memory leaks and network latency, but their process-level analysis, instead of thread-level, lacks granularity for pinpointing degradation causes. Their approach targets system-wide issues, capturing co-located interference but overlooking application-specific constraints (e.g., IPC, lock, and scheduling contention).

\paragraph{\textbf{Instrumentation in Performance Degradation Deconstruction}} In the event of application KPI degradation, root-cause identification techniques are crucial for identifying performance issues \cite{wyatt2020canario}. 
BARO~\cite{pham2024baro} and N-Sigma~\cite{li2022causal,lin2018microscope} utilize statistical methods to detect distribution shifts in time-series metrics before and after anomalies, flagging them as potential root causes. However, these methods are overwhelmed by large datasets because they test full dataset instead of a smaller metric subset based on application dependency analysis (e.g., selective thread tracking). For multi-tier applications, previous research~\cite{cohen2004correlating, zhang2005ensembles, gan2021sage, gan2019seer} use hardware resource usage statistics to infer root causes, while systems like FIRM~\cite{qiu2020firm} employ support vector machines to find critical paths of a microservice dependency graph and localise performance variability. However, \cite{sriraman2018mutune} emphasizes that application thread models significantly impact performance, necessitating RCA at this granularity for finer-grained control.

To mitigate the granularity concerns of the above approaches, Wang et al.~\cite{wang2020recorder}, and Ather et al.~\cite{ather2024drilling} perform a more detailed analysis by collecting IO tracing data through multi-layer IO stack instrumentation to deduce the cause of degradation. However, their narrow IO stack focus limits diagnostic capabilities. Seo et al. \cite{seo2022nosql} propose a more generalisable approach for diagnosing NoSQL database performance issues, tracing application-kernel interactions via system call instrumentation. We adopt a similar approach, tracing specific interactions between application threads and the kernel. However, due to our approach's broader applicability, we differ in the metrics we devise and our collection methods.

\section{Instrumentation Design and Architecture}\label{sec:archi}
Application monitoring typically requires domain-specific knowledge to identify metrics that reveal causes of performance degradation. These often manifest as application-level metrics curated by experienced users and developers. However, these metrics' specificity to particular applications means that introducing new application types necessitates additional domain knowledge to interpret metric importance, thus increasing complexity. While these metrics aid in debugging application activity, they offer limited guidance on mitigating QoS degradation resulting from resource interference.

To mitigate such challenges, system-level metrics are frequently used to design a generalisable approach for monitoring diverse application types. Following this method, \cite{grohmann2019monitorless, cheng2024fedge, yanggratoke2015predicting, ahmed2018automated} collect readily available system metrics that encode interference in kernel metrics exposed by the \texttt{proc} filesystem. Nevertheless, numerous situations demand a more fine-grained set of metrics to detect specific degradation scenarios, such as: when a crucial thread is blocked waiting to receive data from a socket, where per-socket thread statistics can reveal the source of the application's QoS degradation. To this end, this section primarily focuses on ``\textit{what metrics should be collected for application-agnostic performance degradation diagnosis?}''.

\begin{figure}[t]
    \centering
    \includegraphics[width=\columnwidth]{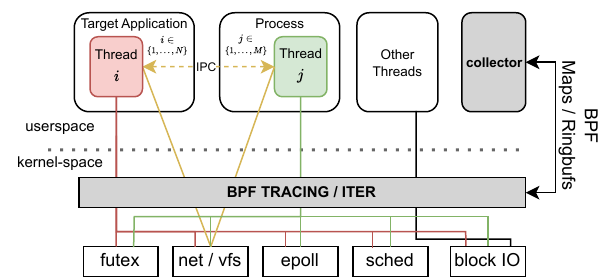}
    \caption{{\textbf{Instrumentation architecture:} 
    Illustration of the required components to monitor a target application. Our instrumentation also accounts for IPC and therefore thread groups that have communicated with the target application will also be monitored.}} \upp\upp\upp
    \label{fig:architecture}
\end{figure}

\begin{table*}[!t]
\centering
    \caption{Enumeration of the instrumented metrics. We present the collected granularity and scope for each metric derived from the most \emph{time}- and \emph{frequency}-sensitive system calls~\cite{wang2023characterizing} and five core system resources summarized by ~\cite{sites2021understanding}.}
\label{table:metrics}
\vspace*{-0.1cm}
\setlength{\tabcolsep}{0.5pt}
\resizebox{\linewidth}{!}
{
\begin{tabular}{|c|c|c|c|c|c|}
\cline{1-6}
\hline
    \textbf{System Calls}&\textbf{Subsystems} & \textbf{Metric} & \textbf{Scope} & \textbf{Granularity} & \textbf{Metric description}
\\
\cline{1-6}
\hline
\hline

        \multirow{5}{*}{\shortstack{fork}} & \multirow{5}{*}{Scheduler} & \emph{runtime} & target & thread & time spent on-CPU \\
        \cline{3-6}
        & & \emph{rq\_time}  & target & thread & time waiting on runqueue \\
        \cline{3-6}
        & & \emph{block\_time}  & target & thread & time in uninterruptible sleep \\
        \cline{3-6}
        & & \emph{iowait\_time}  & target & thread & time in uninterruptible sleep and in \emph{iowait}  \\
        \cline{3-6}
        & & \emph{sleep\_time}   & target & thread & time spent in interruptible sleep \\
        \cline{3-6}
        \hline
        \hline
        
        \multirow{5}{*}{\shortstack{open, close, stat,\\ read, write, sendto,\\ recvfrom, select, poll}} & \multirow{5}{*}{\shortstack{VFS, Network,\\IO Multiplexing,\\Block IO}} & \emph{pipe\_wait\_time} & target & thread$\rightarrow$pipe &  time thread waits for pipe \\
        \cline{3-6}
        & & \emph{pipe\_wait\_count} &  target & thread$\rightarrow$pipe & thread pipe wait frequency \\
        \cline{3-6}
        & & \emph{socket\_wait\_time}& target & thread$\rightarrow$socket & time thread waits for socket \\
        \cline{3-6}
        & &\emph{socket\_wait\_count}& target & thread$\rightarrow$socket & thread socket wait frequency \\
        \cline{3-6}
        & & \emph{sector\_count} & global & thread$\rightarrow$device & thread sector requests to device \\
        \cline{3-6}
        \hline
         \hline
       \multirow{3}{*}{\shortstack{epoll\_create, epoll\_ctl,\\ epoll\_wait}}  & \multirow{3}{*}{\shortstack{Multiplexing IO}} & \emph{epoll\_wait\_time} & target & thread$\rightarrow$epoll & thread epoll wait time \\
       \cline{3-6}
        & & \emph{epoll\_wait\_count} & target & thread$\rightarrow$epoll & thread epoll wait frequency \\
        \cline{3-6}
        & & \emph{epoll\_file\_wait} & target & epoll$\rightarrow$virtual file & time epoll waits for file \\
        \cline{3-6}
        \hline
         \hline
         \multirow{3}{*}{\shortstack{futex}} & \multirow{3}{*}{\shortstack{futex}} & \emph{futex\_wait\_time} & target & thread$\rightarrow$futex & time waiting for a futex \\
         \cline{3-6}
        & & \emph{futex\_wait\_count} & target & thread$\rightarrow$futex & futex wait frequency \\
        \cline{3-6}
        & & \emph{futex\_wake\_count} & target & thread$\rightarrow$futex & futex wake frequency \\
        \cline{3-6}
        \hline
\end{tabular} 
}\upp\upp\upp
\end{table*}

\paragraph{\textbf{Instrumentation Design}} Motivated by the above observations, our instrumentation design goals are: \circled{\bf 1} To offer a set of metrics applicable to widely used application types (such as online data-intensive applications); \circled{\bf 2} To gather metrics at a granularity that offers an effective trade-off between diagnosability and overhead; and \circled{\bf 3} To operate without source code instrumentation.

Considering vast number of kernel probes (approximately $80471$\cprotect\footnote{This value was obtained on a machine running \textbf{linux 6.8.12} with \verb|bpftrace -l "kprobe:*"| \texttt{|} \verb|wc -l|.}) available at different granularity levels, a key concern when instrumenting is deciding \emph{what} to monitor. 
Aligned with our design goals, we target kernel functionalities critical to application execution time and frequency, referencing Wang et al.'s~\cite{wang2023characterizing} analysis of time- and frequency-sensitive system calls in popular \emph{Dockerhub} applications under diverse workloads, and the five core system resources summarized by ~\cite{sites2021understanding}.

Table \ref{table:metrics} shows the set of 16 eBPF-based instrumentation  metrics, derived from analysing time-consuming kernel resources used by typical OLDI applications (see Section~\ref{sec:methodology} for metric instrumentation methodology details). Each metric offers visibility into a specific kernel subsystem, with granularity based on observability needs to identify common performance issues like disk~\cite{neuwirth2017automatic} and lock contention~\cite{liu2020mitigating}. 

A \textit{subsystem} refers to the specialised kernel component that is traced to collect the metric. We collect metrics for applications interacting with scheduling, virtual file system (VFS), networking, futex, multiplexing IO, and blockIO Linux subsystems~\cite{gregg2014systems}, corresponding to frequently used and time-sensitive system calls~\cite{wang2023characterizing}. For example, \emph{epoll}, \emph{select}, and \emph{poll} relate to multiplexing IO subsystem, while \emph{sendto} and \emph{recvfrom} system calls pertain to network subsystem. 
Our focus on the futex subsystem  stems from Sites' et al.~\cite{sites2021understanding} fifth fundamental resource, which emphasizes tracing shared data access in multithreaded programs due to possible interference.

For each instrumented metric, we identify the target subsystem and the desired \textit{Granularity} of performance degradation identification. For instance, \textbf{``thread$\rightarrow$futex''} \textit{Granularity} indicates, \emph{futex\_wait\_\{time,count\}} metrics are collected per thread for a specific futex resource. Finally, a \textbf{``target''} \textit{Scope} indicates a metric is only collected for threads that belong to processes that are part of, or communicated with, the target application, whereas a \textbf{``global''} \textit{Scope} refers to metrics that are collected for all system threads. 

\paragraph{\textbf{Instrumentation Architecture}} Figure~\ref{fig:architecture} illustrates our instrumentation architecture. When a target application executes probed kernel functions, custom \emph{ebpf} programs compute raw statistics about thread interactions with the six kernel subsystems. Statistics are shared with a userspace component (\textit{collector}) via maps and ring buffers, which then transforms the raw data into the metrics shown in Table~\ref{table:metrics}. To minimize resource usage, only statistical summaries are \emph{read every second}, despite high-frequency event tracing by the \emph{ebpf} programs. Next, we describe the probes and methods used to realise the aforementioned metrics.

\section{Instrumentation Methodology}\label{sec:methodology}
This section describes our instrumentation methodology, i.\,e., \emph{how} to collect the metrics in Table~\ref{table:metrics}. These metrics are inferred from kernel statistics aggregated \emph{every second}, including CPU statistics from the scheduling subsystem; pipes, sockets, and futexes for inter-thread/process communication; multiplexing IO for asynchronous programming and high-performance web-servers; and disk IO for applications with persistence guarantees.

\paragraph{\textbf{Scheduling}} A thread's scheduling metrics refers to its time spent in CPU-related states, which are categorized into three main types. \emph{Running} state describes when a thread executes on a CPU core, corresponding to our \emph{runtime} metric in Table \ref{table:metrics}. In the \emph{Idle} state, the thread is not executing; external activity must place it back on a CPU core's runqueue. This state can be further divided into interruptible and uninterruptible states, where the time spent in \verb|TASK_INTERRUPTIBLE| state maps to the \emph{sleep\_time} metric, while the time in the \verb|TASK_UNINTERRUPTIBLE| state corresponds to our \emph{block\_time} metric. A subset of this uninterruptible time is captured by our \emph{iowait\_time} metric, which reflects the time spent in the \verb|TASK_UNINTERRUPTIBLE| state when the \verb|in_iowait| field is set to \verb|1|. Finally, in the \emph{Runnable} state, a thread is ready to run on a core but will not execute while another thread is active. Preemptive schedulers, such as Linux's completely fair scheduler (CFS), manage core time distribution among threads in a runqueue, ensuring fair allocation. The time a thread spends in runqueue without being scheduled is recorded by our \emph{rq\_time} metric.

\paragraph{\textbf{Pipes and Sockets}} In multi-threaded architectures it is common to segregate functionality between the different threads belonging to an application. Sending and receiving data to and from other threads is a fundamental requirement of these architectures to coordinate work. Pipes and sockets are two such mechanisms that facilitate this communication.

When a thread writes to a pipe, data accumulates in a kernel memory buffer. This data is then forwarded in a First-In First-Out (FIFO) fashion to any thread that reads from the same pipe. However, userspace threads use file descriptors (\verb|fd|) to identify the read and write ends of a pipe. File descriptors take the form of integer values that are used to extract the underlying \verb|struct file| pointed at by this descriptor. Multiple file descriptors can point to the same backing resource with varying read/write permissions. For example, there may exist a file descriptor with write permission, and another with read permission, both referring to the same pipe. Therefore, to uniquely identify the backing resource pointed at by the file descriptor, we create a \emph{Backing Resource Identifier} (\emph{BRI}) with a combination of the inode's virtual file system identifiers contained in the \verb|struct file| data structure, namely: \verb|i_ino| representing the \emph{inode identifier}; and \verb|s_dev| denoting superblock device identifier. In the above example with both file descriptors writing to same pipe, the pipe's \verb|i_ino| and \verb|s_dev| will be equivalent, resulting in same \emph{BRI}.

To instrument the time and frequency a thread waits in read/write calls to a FIFO file, given by \emph{pipe\_wait\_\{time,count\}} metrics (see Table~\ref{table:metrics}), we trace the kernel's \verb|vfs_{write,read}| functions. The \verb|struct file| argument in these functions is used to compute the \emph{BRI}, enabling attribution of a thread's wait time to specific pipes. For example, if the \verb|O_NONBLOCK| flag is unset (i.\,e., changing the file descriptor's behavior from non-blocking to blocking mode), our metrics (i.\,e., \emph{pipe\_wait\_\{time,count\}}) account for the time a thread sleeps in its read call while waiting for data to be written on the other end of the pipe.

Sockets, unlike pipes, operate as two-way communication streams with both transmit/receive buffers. For inter-thread communication via sockets, each thread must have a socket mapped to the other, determined by the socket family, type, and protocol. For example, \verb|AF_INET|, \verb|SOCK_STREAM|, and \verb|IPPROTO_TCP| sockets communicate via a TCP connection identified by the protocol 5-tuple:
  $\langle \mathtt{TCP\,protocol},\ \mathtt{source\,IP},\ \mathtt{source\,port},\ \mathtt{destination\,IP},\\ \mathtt{destination\,port} \rangle$.

To monitor the time and frequency a thread waits for socket data reception, denoted by \emph{socket\_wait\_time} and \emph{socket\_wait\_count} metrics (see Table \ref{table:metrics}), we probe kernel's \verb|sock_recvmsg| function. When invoking this function, the kernel passes socket specific data that identifies its respective domain specific source and destination identifiers. We currently support \verb|AF_INET|, \verb|AF_INET6|, and \verb|AF_UNIX| domains. Similarly, on the sending end, we probe domain-specific functions for \verb|AF_INET|, \verb|AF_INET6|, and \verb|AF_UNIX|, i.\,e., \verb|inet_sendmsg|, \verb|inet6_sendmsg|, and \verb|unix*sendmsg| to monitor \emph{socket\_wait\_\{time,count\}} metrics of thread socket data transmission. By collecting these metrics, we can estimate the duration a thread is suspended during interactions with pipe/socket subsystems, highlighting straggling external dependencies that may impede an application's progress.

\paragraph{\textbf{Futex}} Fast userspace mutexes (or futexes) serve as a thread synchronisation mechanism provided by the kernel via the futex system call~\cite{drepper2005futexes}. Two threads synchronize by agreeing on a memory address (\verb|uaddr| parameter) referencing an integer value. For thread sleep requests, the kernel compares the value at \verb|uaddr| with the \verb|val| parameter; if equal, the thread sleeps. To wake sleeping threads, another thread calls futex with the same \verb|uaddr|, specifying \verb|val| threads to wake. Common \verb|val| values are \verb|1| (i.\,e., wake one thread) and \verb|INT_MAX| (i.\,e., wake all threads)~\cite{drepper2005futexes}.

The futex subsystem enables two key use cases. First, it can mediate resource access, limiting the number of threads that can access a particular resource. This functionality underpins various locking mechanisms such as read-write locks, mutexes, and semaphores. Second, futexes can be used as a work scheduling mechanism. In this scenario, a thread is put to sleep when it has no work to process and is awakened by another thread when work becomes available. An example of this is seen in Rust's standard library, which uses futexes in its multi-producer single-consumer (mpsc) message passing implementation. Here, the consumer is parked using futex~\cite{rust_parker} when the message queue is empty and is awakened by a producer when a new message is appended.

To monitor thread interactions with the futex subsystem, we instrument the \verb|sys_enter_futex| (i.\,e., thread entering the futex system call) and \verb|sys_exit_futex| (i.\,e., thread exiting the futex system call) tracepoints. We use the \verb|op| argument to distinguish between futex wake and sleep operations, collecting: total time and frequency a thread sleeps on a specific \verb|uaddr|, represented by \emph{futex\_wait\_time} and \emph{futex\_wait\_count} metrics (see Table~\ref{table:metrics}). The frequency of successful wake operations on a \verb|uaddr|, i.e., waking at least one thread, is given by \emph{futex\_wake\_count} metric. This enables observing an application's lock contention and work scheduling patterns, crucial for understanding how degraded performance propagates across different application threads.

\paragraph{\textbf{Multiplexing IO}} In synchronous programming, with the \verb|O_NONBLOCK| flag unset, a thread executes a system call (e.g., recv) on a single file descriptor (e.g., socket) and must wait for its completion before proceeding to other tasks. High-performance web servers challenged this approach by introducing a scenario where a single thread manages multiple connections concurrently, waiting for incoming data on multiple connections simultaneously \cite{brecht2006evaluating}. For network IO-intensive servers, this approach offers several advantages, including: enhanced CPU cache utilization; decreased memory footprint; and minimized context switching.

The kernel mechanism enabling this new paradigm is multiplexing IO. In essence, a thread specifies which events it wants to consume from a particular file descriptor and adds it to an interest list. It can then wait for all registered \verb|fd|s using a specific system call, which returns when one or more \verb|fd|s have events ready for consumption
\footnote{There are other cases that may induce an early return, such as, signal handling and elapsed timers}. 
Linux offers three APIs for this purpose: select, poll, and epoll~\cite{kerrisk2010linux}.

Select and poll share a similar interface where a single system call is used to both express a thread's interests and wait until an event is ready for consumption from the file descriptor. For select, we probe the entry and exit of \verb|do_select|, while for poll, we probe the entry and exit of \verb|do_sys_poll|. Upon returning from these functions, we collect the \emph{BRI} of all registered file descriptors and increment each \emph{BRI}'s wait time (e.\,g., \emph{pipe\_wait\_time} and \emph{socket\_wait\_time} metrics in Table \ref{table:metrics}). In essence, a single call to select or poll attributes a thread's wait time to all registered \emph{BRI}s (rather than just one).

In contrast, epoll's API offers three system calls for interacting with the kernel's event poll subsystem. First, \verb|epoll_create| generates an epoll resource and returns a file descriptor (\verb|epfd|) for userspace code to manage its interest list. Second, a thread calls \verb|epoll_ctl| with the \verb|epfd| to add/remove file descriptors from the interest list. Third, a thread invokes \verb|epoll_wait| with the \verb|epfd| to wait for its registered \verb|fd|s.

Given epoll's use of separate system calls to register interest in a \verb|fd|'s events (i.\,e., using \verb|epoll_ctl|) and to wait for registered \verb|fd|s (i.\,e., using \verb|epoll_wait|), our eBPF programs keep a copy of all \verb|fd| \emph{BRI}s registered with each epoll resource. Updating this copy is done, when \verb|fd| events are registered and de-registered using \verb|ep_remove| and \verb|ep_insert| functions.

Furthermore, given that an \verb|epfd| is used by userspace threads to reference an epoll resource, an epoll's \emph{BRI} is derived from the kernel memory address pointing to its \verb|struct eventpoll| data structure. This approach differs from the previous method of calculating a file descriptor's \emph{BRI} because epoll \emph{inodes} are part of the anonymous \emph{inode} filesystem, which reuses the same \emph{inode} for all resources of the same type. Given that multiple threads can wait for the same \verb|epfd|, our first epoll related metric measures the time and frequency a thread spends waiting for a particular epoll \emph{BRI}, represented by \emph{epoll\_wait\_time} and \emph{epoll\_wait\_count} metrics (see Table~\ref{table:metrics}).  Additionally, we account for the time a registered file descriptor is awaited for during an \verb|epoll_wait| system call (i.\,e., \emph{epoll\_file\_wait} metric), attributed to the epoll \emph{BRI} it was registered with.

\paragraph{\textbf{Disk IO}} Given the data persistence requirements of many OLDI applications (e.g., databases and message brokers), their high-level performance metrics can be significantly affected by the maximum throughput achieved when writing data to disk. This is typically because these applications can only confirm message receipt after successful data storage. However, given that disk access is orders of magnitude slower than CPU cache or RAM access, monitoring the time an application's thread spends waiting for disk request completion is crucial. While this information is partially covered by the \emph{iowait\_time} and \emph{block\_time} scheduling metrics, \emph{sector\_count} metric (see Table~\ref{table:metrics}) provides a more comprehensive view of block device activity. It allows for calculating the total in-flight device requests and determining the proportion of these requests belonging to a target application.

\section{Experiments}
\label{sec:experiments}
This section presents experimental setup, selected applications, induced workload and performance degradation scenarios to evaluate diagnostic capability of instrumented metrics.

\paragraph{\textbf{Experimental Setup}} We implemented instrumentation method as a Rust executable that leverages \emph{libbpf} to interface with \emph{bpf}. It cyclically gathers kernel statistics \emph{every second} using our custom eBPF programs, transforms them into metrics listed in Table \ref{table:metrics}, and persists them for further analysis. Instrumented metrics' complete source code and reproducibility artifact is available in an \emph{online} \verb|GitHub| repository~\cite{Prism}. We executed our experiments on \verb|i7-11850H Intel| server @ \SI{2.50}{\giga\hertz} running  Ubuntu 22.04 LTS, Jammy Jellyfish (\verb|x86_64|) OS with \verb|6.8.12| Linux kernel, $8$ CPU cores, $2$$\times$\SI{16}{\giga\byte} \verb|DDR4| RAM, and \SI{1}{\tera\byte} SSD disk.

\paragraph{\textbf{Selected Applications}} We selected \emph{seven} online data-intensive applications for validation. Our application set includes a relational and a NoSQL database, \emph{MySQl} and \emph{Apache Cassandra}. We also selected \emph{Redis} which fits into the category of in-memory Key-value datastores, while \emph{Kafka} is a message broker middleware. To incorporate a broader spectrum of applications, our evaluation set also includes an open-source search service \emph{Apache Solr} and a microservice-based benchmark application \emph{TeaStore}~\cite{von2018teastore}. Finally, we also evaluate a machine learning inference (ML-inference) service wrapped around a sentiment analysis model~\cite{perez2021pysentimiento} in a \emph{FastAPI web-app}, deployed using a \verb|Python-based| webserver gateway interface \emph{gunicorn}.

Most of these selected applications span across a wide range of domains and have been widely used in previous literature\cite{grohmann2019monitorless, xing2023hybrid, palit2016demystifying, wang2023characterizing}. Additionally, the chosen applications provide the variability with regards to combining and utilizing different threading models, inter-process communication (IPC) methods, and user-space virtual machines. For instance, \emph{MySQl} leverages a thread pool to handle its connections whereas \emph{Kafka} employs multiplexing IO to handle the different client requests. Concerning userspace virtual machines, \emph{Kafka}, \emph{Cassandra} and \emph{Solr} operate on the Java Virtual Machine (JVM), while our ML-inference service runs on the Python Virtual Machine. Lastly, with regards to IPC, \emph{Kafka} and \emph{Teastore} leverage pipes to communicate between threads, while futex mechanisms are used by all applications for synchronisation and work scheduling activities.

\paragraph{\textbf{Workload Generation}} We used YCSB~\cite{cooper2010benchmarking} and TPCC~\cite{leutenegger1993modeling} benchmarks to generate read- and update-intensive workloads for \emph{MySQL}, while for \emph{Cassandra}, only YCSB was used. For \emph{Solr}, we utilized the Cloudsuite benchmark~\cite{palit2016demystifying} and modified it to enable variable load patterns. In case of \emph{Redis}, we used the official ~\cite{redis_benchmark} and the memtier~\cite{redis_memtier} benchmarks to saturate the server. Workload for the ML-inference server and \emph{Teastore} applications  was generated using the Locust~\cite{locust} framework. We applied the same workload pattern for both cases, varying load between \SIrange[range-phrase=\,--\,]{2}{60}{} users. For the ML-inference server specifically, the server is queried about the sentiment of tweets we extract from a twitter dataset \cite{twitter_dataset}, whereas \emph{Teastore}'s requests are generated by the request generator provided in their official repository \cite{von2018teastore}. Finally, for \emph{Kafka} we generated a variable workload pattern by creating a custom producer with an increasing production rate of upto \SI{450,000}{} events per second.

\paragraph{\textbf{Degradation Scenarios}} We utilized a combination of two approaches to degrade the selected applications performance. The first approach used benchmarks to increase the application's load to the point of saturation, impairing its performance, similar to ~\cite{gregg2013thinking}. The second approach induced a combination of CPU, disk and lock contention to interfere with the application's activity and hinder its performance. CPU contention involved either underprovisioning CPU resources (e.g. \emph{Teastore}, \emph{Solr}, and \emph{Cassandra}) or competing for same core on-cpu time (e.g. \emph{Redis}). As for disk contention, inspired by stress-ng's \cite{stressng} \verb|hdd| stress procedure, we start 80 threads that execute synchronous writes to disk (\verb|O_SYNC| flag set) at their maximum achievable rate, to compete for the underlying disk resource. Lastly, with regards to lock contention, we induced this type of contention on MySQL due to its ACID compliance \cite{mysql_acid}. We performed this by executing two distinct YCSB workloads reading from and writing to the same dataset.

\paragraph{\textbf{Target Performance Metrics}} To evaluate the instrumented metrics' robustness to different high-level KPIs, we vary the target performance metrics in each experiment. In case of \emph{MySQL}, we use two target metrics: $95^{th}$ percentile \emph{response time} (seconds) and \emph{throughput} (requests per second), which measures \emph{MySQL's} performance as seen by the YCSB and TPCC workloads. \emph{Cassandra's} performance is measured using median \emph{response time}, while \emph{Kafka} uses the total production rate (i.e. \emph{throughput}) in events per second, as a proxy of its performance. Similarly, we used $95^{th}$ percentile \emph{response time} as a target metric to estimate \emph{Solr}, ML-inference server, \emph{Redis} and \emph{Teastore} application performance.

\section{Results}\label{sec:results}
This section illustrates the use of implemented \emph{eBPF-based} metrics from Table~\ref{table:metrics} to analyze and identify root causes of performance degradation in applications under the workload patterns and scenarios from Section~\ref{sec:experiments}. We reference real kernel identifiers for reproducibility and include only the most relevant metrics for each scenario.

\paragraph{\textbf{Selective Thread Tracking}} Our root cause identification approach is consistent across all applications and scenarios, using metrics in Table~\ref{table:metrics}. Initially, we identify a set of entrypoint threads that present IPv4/IPv6 socket communication using \emph{socket\_wait\_\{time,count\}} metrics, which are added to a \emph{thread tracking list} for degradation analysis. Raw and transformed metric timeseries for identified threads are examined, flagged as potential root cause candidates if their distribution shift matches target metric's observed shift. For flagged IPC-related metrics (i.e., \emph{futex*}, \emph{socket*}, \emph{pipe*} metrics), metric granularity allows identification of counterpart threads involved in communication through same \emph{BRI}, suggesting thread dependencies. Counterpart threads are also added to \emph{thread tracking list} until no unique thread remains. The granular metric collection simplifies root cause identification by narrowing search to inter-dependent threads and tracing a clear path from constrained to outward-facing threads.

Next, we summarize the degradation analysis results for each application in Section~\ref{sec:experiments}, grouped by expected primary resource constraint. While scenario knowledge aids predictability, our goal is to demonstrate that the same conclusions are reachable using only the target KPIs and metrics from Table~\ref{table:metrics}.

\begin{figure}[t]
    \centering
    \upp
    \subfloat[\label{fig:mysql:throughput}]{
      \includegraphics[width=0.31\columnwidth]{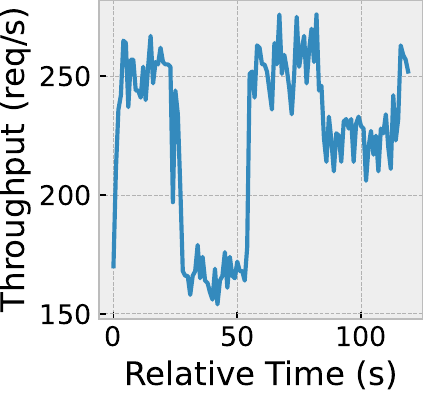}
    }
    \subfloat[\label{fig:mysql:latency}]{
      \includegraphics[width=0.31\columnwidth]{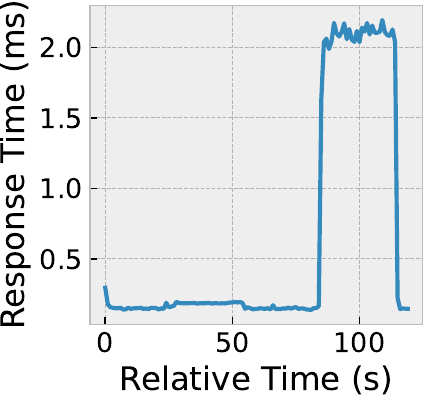}
    }
    \subfloat[\label{fig:mysql:device_reqs}]{
      \includegraphics[width=0.32\columnwidth]{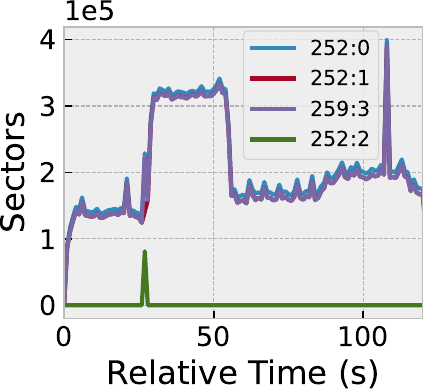}
    }\\  \upp\upp 
    \subfloat[\label{fig:mysql:device_share}]{
      \includegraphics[width=0.31\columnwidth]{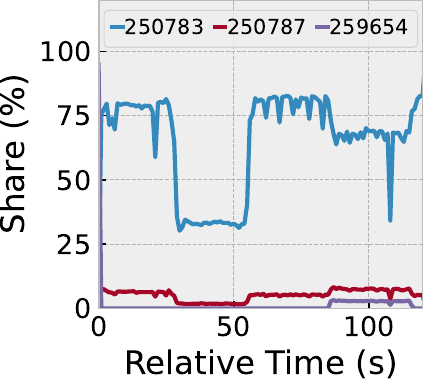}
    }
    \subfloat[\label{fig:mysql:lock_futex_wait}]{
    \upp\upp
      \includegraphics[width=0.31\columnwidth]{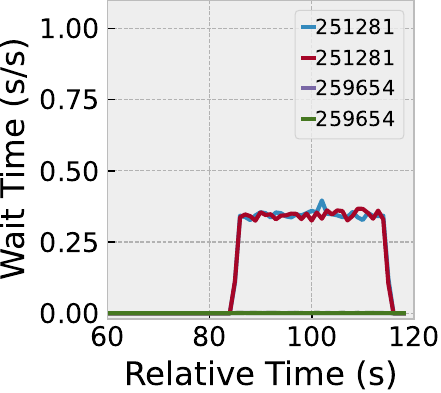}
    }
    \subfloat[\label{fig:mysql:lock_futex_wake}]{
    \upp\upp
      \includegraphics[width=0.30\columnwidth]{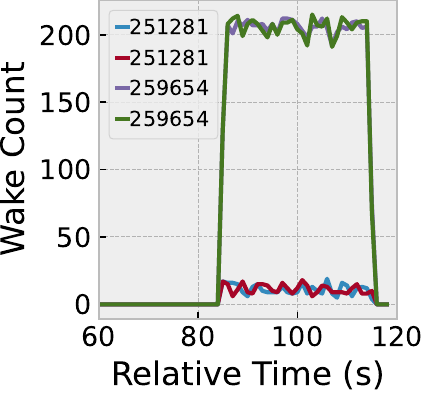}
    }\\ \upp\upp 
    \subfloat[\label{fig:mysql:disk_sched_iowait}]{
    \upp\upp
      \includegraphics[width=0.31\columnwidth]{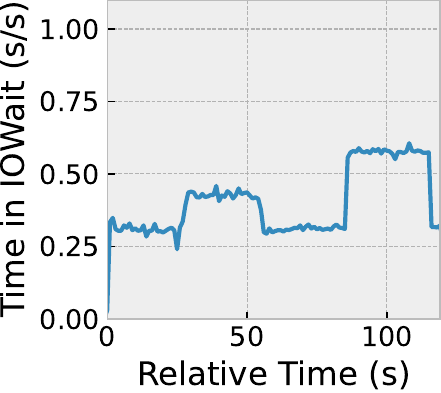}
    }
    \subfloat[\label{fig:mysql:disk_futex_wait}]{
    \upp\upp
      \includegraphics[width=0.31\columnwidth]{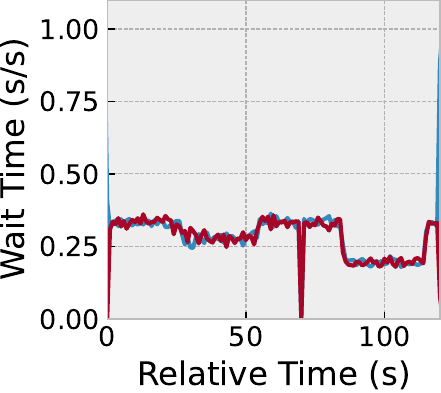}
    }
    \subfloat[\label{fig:mysql:disk_futex_wake}]{
    \upp\upp
      \includegraphics[width=0.30\columnwidth]{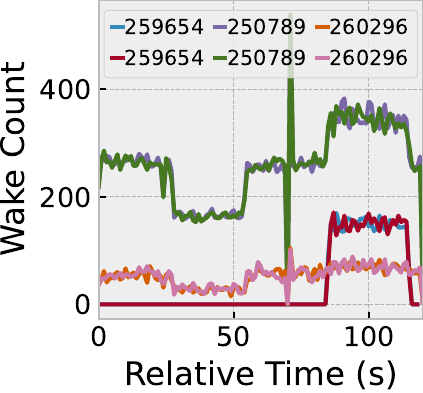}
    }
    \vspace*{-0.1cm}
    \caption{\textbf{MySQL target metric observation and performance degradation deconstruction:} 
    (a) TPCC workload throughput; (b) YCSB read-intensive workload $95^{th}$ percentile response time; (c) Sector requests per device (identified by their \texttt{major:minor|values}; (d) Device request share for MySQL threads; 
    (e) Per thread futexes' (\texttt{0x76d594012f30, 0x76d594012f34}) wait time; (f) Per thread futexes' (\texttt{0x76d594012f30, 0x76d594012f34}) wake activity; (g) Thread \texttt{250787}'s \emph{iowait\_time}; (h) Thread \texttt{250787}'s futexes' (\texttt{0x76d5fc83c994, 0x76d5fc83c990}) wait time; (i) Per thread futexes' (\texttt{0x76d5fc83c994, 0x76d5fc83c990}) wake activity.
    } \upp\upp\upp
\end{figure}

\subsection{Disk Constrained}
\paragraph{\textbf{MySQL}}\textbf{Experiment Scenario \& Target Metric Observation.} For the \emph{MySQL} experiment, we executed TPCC and YCSB read-intensive workloads for \SI{120}{\second}. We implemented two interventions during this period. The first, a disk contention workload, began \SI{25}{\second} into the execution and continued for \SI{30}{\second}. The second intervention, starting 85 seconds into the execution and lasting \SI{30}{\second}, involved executing a YCSB update-intensive workload inducing lock contention with the YCSB read-intensive workload, and disk contention with the TPCC benchmark due to its update operations.

Figures~\ref{fig:mysql:throughput} and~\ref{fig:mysql:latency} shows TPCC's throughput and YCSB's $95^{th}$ percentile latency throughout the entire execution. We can observe that the first intervention significantly reduces the server's TPCC benchmark throughput performance but does not affect the YCSB read-intensive workload. As expected, TPCC heavily relies on disk interactions, while YCSB primarily reads from RAM. In contrast, the second intervention, which competes for the same YCSB dataset and requires disk writes, degrades both target metrics.

\textbf{Deconstructing Performance Degradation.} For the first scenario, Figures \ref{fig:mysql:device_reqs} and \ref{fig:mysql:device_share} show disk activity via in-flight sector requests per device and device share per thread, derived from the \textit{sector\_count} metric. We observed that device requests increase during the first intervention, with MySQL's thread share decreasing from 80\% to 34\%.

For the second intervention, we investigate two degradation scenarios: lock contention between YCSB read- and update-intensive benchmarks; and disk contention between the YCSB update-intensive and TPCC workloads. With regards to lock contention, Figure \ref{fig:mysql:lock_futex_wait} shows thread \verb|251281|'s \textit{futex\_wait\_time} increasing from \SIrange[range-phrase=\,--\,]{0}{0.7}{\second} per second. Given per thread \textit{futex\_wake\_count} for the same futexes presented in Figure \ref{fig:mysql:lock_futex_wake}, this mutual wake activity suggests these futexes control access to a shared resource. This is the activity that led to MySQL's degraded performance with regards to the YCSB read-intensive benchmark considering both threads handle client connections, and that thread \verb|259654| is the thread handling connections for the YCSB update-intensive benchmark.

Concerning disk contention between YCSB update-intensive and TPCC benchmarks, Figure \ref{fig:mysql:disk_sched_iowait} shows thread \verb|250787|'s \emph{iowait\_time} correlating with degradation observed on TPCC's throughput. Figure \ref{fig:mysql:disk_futex_wait} displays thread's decreased \textit{futex\_wait\_time} during second scenario, while Figure \ref{fig:mysql:disk_futex_wake} reveals increased wake activity on the same pair of futexes. These figures highlight the work scheduling role these futexes have for thread \verb|250787|. Furthermore, given that threads \verb|259654| and \verb|260296| handle MySQL connections, the activity shown in Figure \ref{fig:mysql:disk_futex_wake} shows TPCC's stable request rate and YCSB's sudden request surge throughout second intervention. This indicates thread \verb|259654| handles YCSB update-intensive requests, whereas thread \verb|260296| handles TPCC requests.

\begin{figure}[t]
    \centering
    \upp\upp\upp
    \subfloat[\label{fig:kafka:throughput}]{
      \includegraphics[width=0.31\columnwidth]{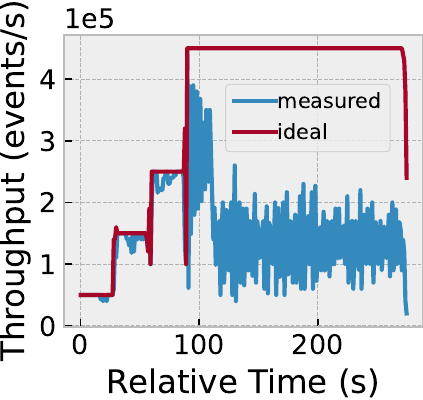}
    }
    \subfloat[\label{fig:kafka:runtime}]{
      \includegraphics[width=0.31\columnwidth]{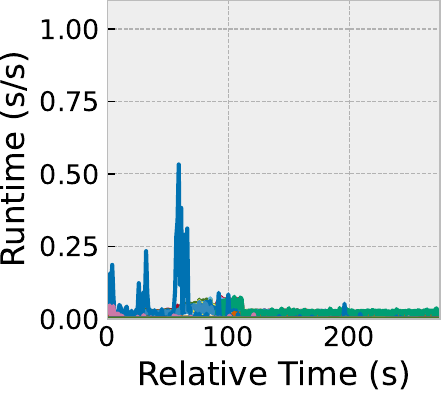}
    }
    \subfloat[\label{fig:kafka:block_time}]{
      \includegraphics[width=0.31\columnwidth]{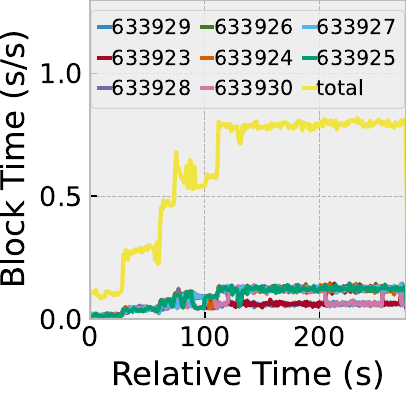}
    }\\ \upp\upp
    \subfloat[\label{fig:kafka:futex_wait}]{
     \upp\upp
      \includegraphics[width=0.31\columnwidth]{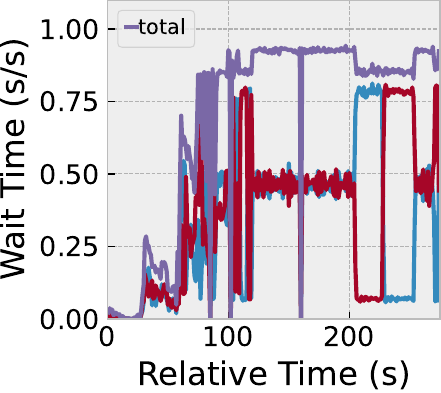}
    }
    \subfloat[\label{fig:kafka:thread_sched}]{
     \upp\upp
      \includegraphics[width=0.30\columnwidth]{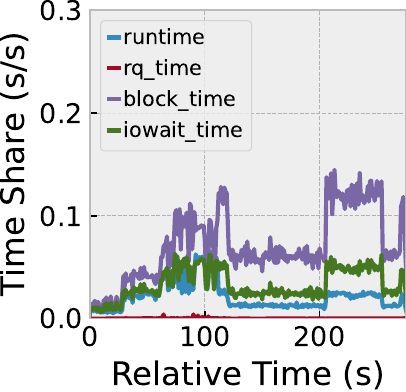}
    }
    \subfloat[\label{fig:kafka:futex_wake}]{
     \upp\upp
      \includegraphics[width=0.31\columnwidth]{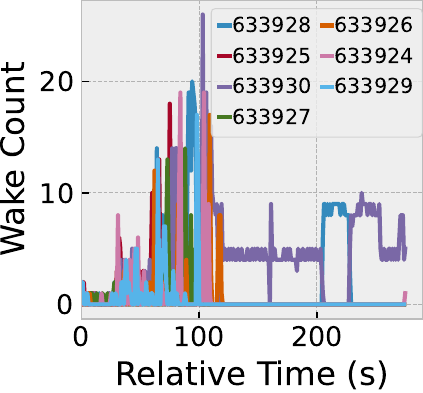}
    }\\ \upp\upp
    \subfloat[\label{fig:kafka:external_wait}]{
     \upp\upp
      \includegraphics[width=0.31\columnwidth]{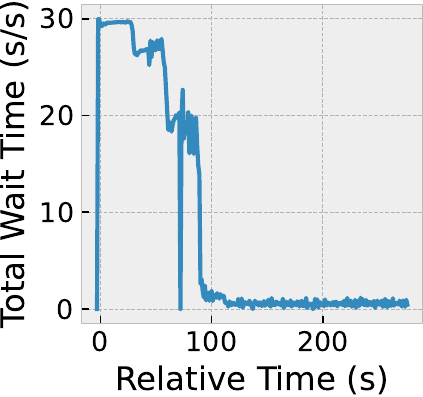}
    }
    \subfloat[\label{fig:kafka:epoll_wait}]{
     \upp\upp
      \includegraphics[width=0.31\columnwidth]{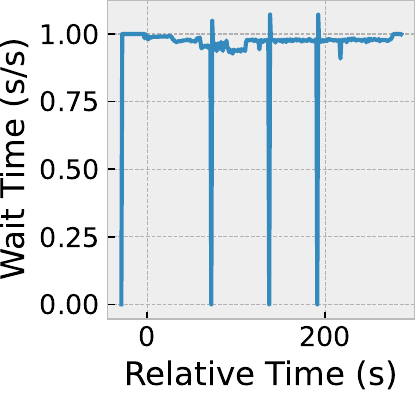}
    }
    \subfloat[\label{fig:kafka:pipe_write}]{
     \upp\upp
      \includegraphics[width=0.32\columnwidth]{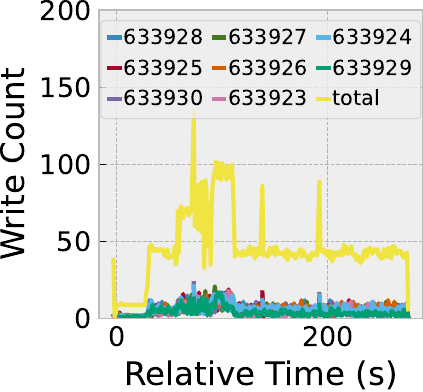}
    }
     \vspace*{-0.2cm}
    \cprotect\caption{\textbf{Kafka target metric observation and performance degradation deconstruction:} (a) Ideal/measured throughput; (b) Kafka threads' \emph{runtime}; (c) Kafka threads' \emph{block\_time}; (d) Thread \verb|633923| futexes \verb|0x74237994a778| and \verb|0x74237994a77c| wait time; (e) Thread \verb|633923| scheduling activity; (f) Wake activity for futex \verb|0x74237994a778|; (g) Combined connection wait time for epoll \verb|ffff9a7a3f5e1740|; (h) Thread \verb|633947| epoll \verb|ffff9a7a3f5e1740| wait time; (i) Pipe \verb|14_2159682| write activity.} \upp\upp\upp\upp
\end{figure}

\paragraph{\textbf{Kafka}}\textbf{Experiment Scenario \& Target Metric Observation.} For Kafka, we increased the load to a point where the broker is unable to handle the total production rate. Figure~\ref{fig:kafka:throughput} compares the increasing ideal throughput with the actual production rate. The system falters at an ideal rate of \SI{450,000}{} events per second, where the observed production rate is $55-77$\% below target.

\textbf{Deconstructing Performance Degradation.} We analyze the application threads' scheduling activities for \emph{Kafka}. Figures \ref{fig:kafka:runtime} and \ref{fig:kafka:block_time} show threads' \emph{runtime} and \emph{block\_time} respectively. Figure \ref{fig:kafka:runtime} displays brief runtime peaks from Java's compiler threads. Figure \ref{fig:kafka:block_time} reveals two key points: total \emph{block\_time} increases with production rate, and only 8 threads (i.\,e., \verb|633923-633930|) out of 94 participate in this activity. Our analysis focuses on this pool of $8$ threads, however we only summarise the behaviour of thread \texttt{633923}, which is generalisable to the remaining threads of the same pool.

Figure \ref{fig:kafka:futex_wait} shows thread \verb|633923|'s \textit{futex\_wait\_time}, oscillating between $85-92$\% during performance degradation. Comparing with Figure \ref{fig:kafka:thread_sched}, \emph{futex\_wait\_time} at 85\% correlates with a peak \textit{block\_time} of \SI{0.12}{\second}, while $92$\% correlates with a lower \emph{block\_time} of \SI{0.06}{\second}, linking futex and block activities. For same futexes thread \verb|633923| waits for, Figure \ref{fig:kafka:futex_wake} displays stable \emph{futex\_wake\_count} activity while \emph{Kafka}'s performance is degraded. Most notably, all threads except the thread under analysis (\verb|633923|) participate in waking these futexes, indicating their use as work schedulers for thread \verb|633923|.

To connect the thread pool activity with the decrease in \emph{throughput} as observed by data producers, we analyze three producer-interacting threads \verb|633947-633949|, with a primary focus on thread \verb|633947|. These threads each have a dedicated epoll \verb|fd|  to await for external producer data, process incoming requests, and return acknowledgments to producers. Figure \ref{fig:kafka:external_wait} shows an aggregation of \verb|epoll_file_wait| for all producer connections, awaited for by epoll resources, with least wait time coinciding with the broker's degraded performance. However, Figure \ref{fig:kafka:epoll_wait} seemingly contradicts this, showing nearly $100$\% \emph{epoll\_wait\_time} for thread \verb|633947|'s epoll backing resource identifier (BRI), as defined in Section~\ref{sec:methodology}. Further analysis reveals a pipe registered with the same epoll resource with decreased write activity (\emph{pipe\_wait\_count}) while the broker was degraded (Figure \ref{fig:kafka:pipe_write}). The pipe-writing threads \verb|633923-633930| are the same threads that were highlighted in our previous analysis as the thread pool with atypical scheduling and futex activity. This indicates the thread pool constrained the server's performance, which led to the reduced \emph{throughput} observed by our target metric.

\begin{figure}[t]
    \centering
    \upp\upp
    \subfloat[\label{fig:cassandra:response_time}]{
      \includegraphics[width=0.31\columnwidth]{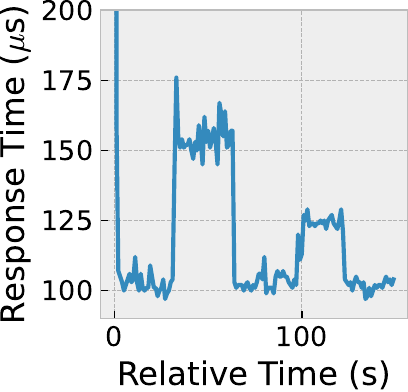}
    }
    \subfloat[\label{fig:cassandra:first_rqtime}]{
      \includegraphics[width=0.31\columnwidth]{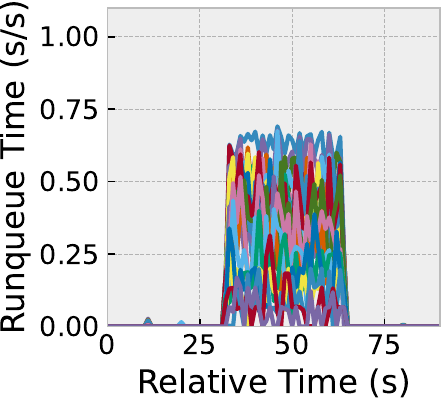}
    }
    \subfloat[\label{fig:cassandra:first_sleep_time}]{
      \includegraphics[width=0.31\columnwidth]{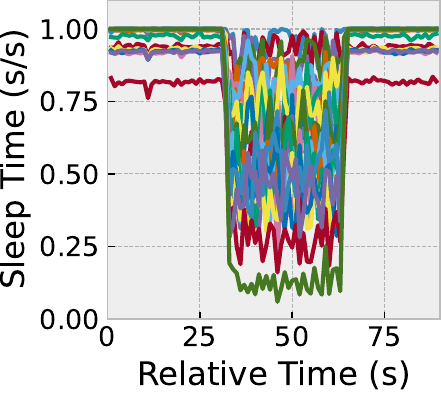}
    }\\ \upp\upp
    \subfloat[\label{fig:cassandra:second_runtime}]{
    \upp\upp
      \includegraphics[width=0.31\columnwidth]{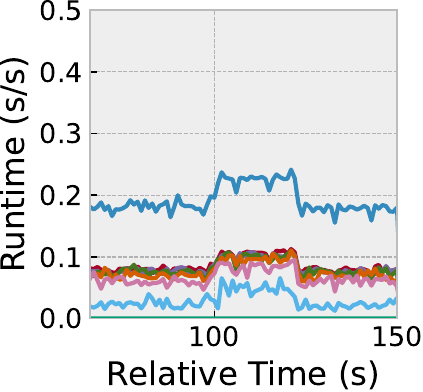}
    }
    \subfloat[\label{fig:cassandra:second_sleep_time}]{
    \upp\upp
      \includegraphics[width=0.31\columnwidth]{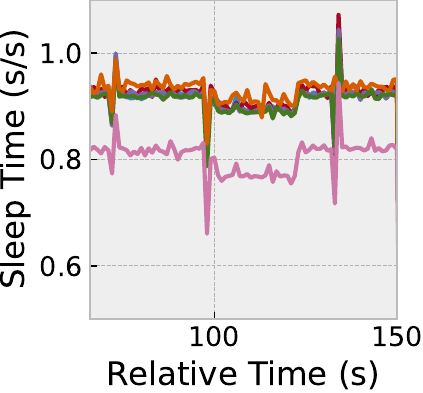}
    }
    \subfloat[\label{fig:cassandra:iowait_time}]{
    \upp\upp
      \includegraphics[width=0.31\columnwidth]{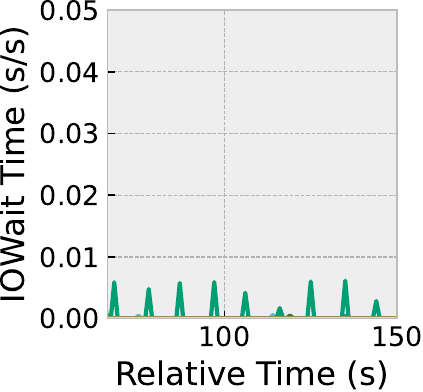}
    }\\ \upp\upp
    \subfloat[\label{fig:cassandra:device_reqs}]{
      \includegraphics[width=0.31\columnwidth]{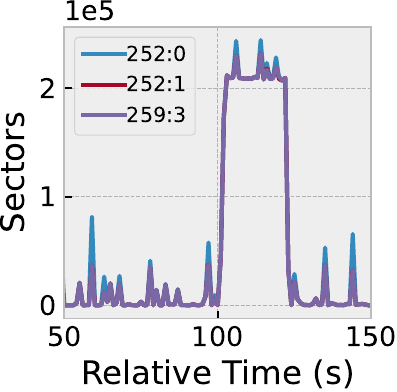}
    }
    \subfloat[\label{fig:cassandra:device_share}]{
      \includegraphics[width=0.31\columnwidth]{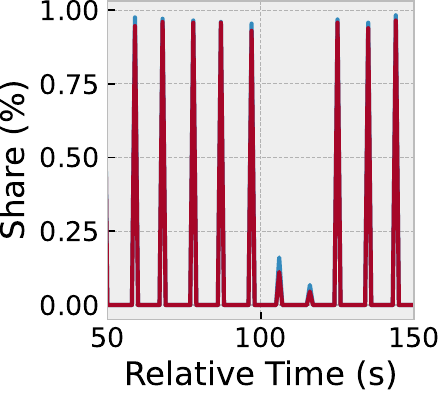}
    }
    \caption{\textbf{Cassandra target metric observation and performance degradation deconstruction:} (a) YCSB update-intensive workload median response time; (b) Cassandra threads' \emph{rq\_time} throughout the first intervention; (c) Threads' \emph{sleep\_time} throughout the first intervention; (d) Threads' \emph{runtime} throughout the second intervention; (e) Threads' \emph{sleep\_time} throughout the second intervention; (f) Thread \emph{iowait\_time} throughout the second intervention; (g) Sector requests per device; (h) Device request share for Cassandra threads.} \upp\upp\upp
\end{figure}

\paragraph{\textbf{Cassandra}}\textbf{Experiment Scenario \& Target Metric Observation.} For Cassandra, we execute a YCSB update-intensive workload, measuring median response time (Figure \ref{fig:cassandra:response_time}) as target metric. Two interventions are implemented: first, in an attempt to induce lock contention, a \SI{30}{\second} YCSB read-intensive workload starts at \SI{33}{\second}, targeting same dataset as YCSB's update-intensive workload; Second, a disk contention workload is introduced between between \SI{100}{\second}-\SI{125}{\second} to interfere with Cassandra disk operations by saturating the device.

\textbf{Deconstructing Performance Degradation.} For the first scenario, Figures \ref{fig:cassandra:first_rqtime} and \ref{fig:cassandra:first_sleep_time} show significant
\emph{rq\_time} increase and, \emph{sleep\_time} decrease, of threads respectively.  Threads depicted in both figures were selected based on absolute Pearson correlation $\geq$ \SI{0.5}{}. Interestingly, $95.8$\% of threads with notable sleep activity also showed interesting \emph{rq\_time} activity. This suggests CPU contention among \emph{Cassandra's} threads causes degraded performance, likely due to insufficient CPU allocation for the container.

In the second intervention scenario, Figure \ref{fig:cassandra:response_time} shows a slight increase in \emph{Cassandra's} median response time. However, Figure \ref{fig:cassandra:second_sleep_time} displays decreased thread sleep time, while Figure \ref{fig:cassandra:second_runtime} shows increased runtime, apparently causing performance degradation. This suggests competing disk workloads led to CPU cache contention, increasing runtime for the same operations (YCSB update-intensive benchmark). This finding was unexpected, as we initially believed \emph{Cassandra's} performance would closely correlate with disk throughput.

To understand the unexpected above disk intervention impact, Figure \ref{fig:cassandra:iowait_time} shows \emph{Cassandra's} \emph{iowait\_time}, suggesting periodic write activity. Figures \ref{fig:cassandra:device_reqs} and \ref{fig:cassandra:device_share} confirm this, displaying sector requests per device and Cassandra threads' request shares, derived from \emph{sector\_count}. \emph{Cassandra's} official documentation~\cite{cassandradocs} reveals that with \texttt{commitlog\_sync} set to \verb|periodic| and \verb|commit_log_sync_period| to \verb|10000|, Cassandra waits \SI{10000}{\milli\second} before \emph{fsync}'ing the commit log, explaining the observed behavior.

\subsection{CPU Constrained}
\begin{figure}[t]
    \centering
    \upp\upp
    \subfloat[\label{fig:solr:users}]{
      \includegraphics[width=0.31\columnwidth]{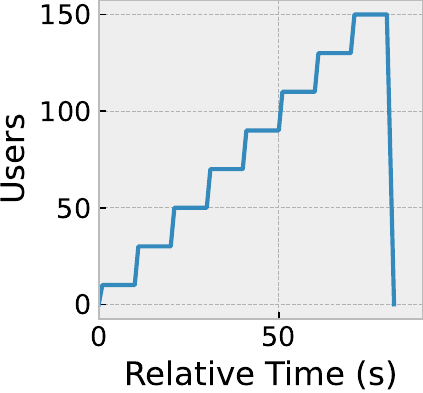}
    }
    \subfloat[\label{fig:solr:target}]{
      \includegraphics[width=0.31\columnwidth]{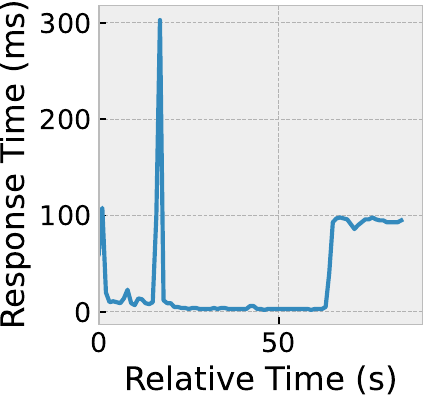}
    }
    \subfloat[\label{fig:solr:rq_time}]{
      \includegraphics[width=0.31\columnwidth]{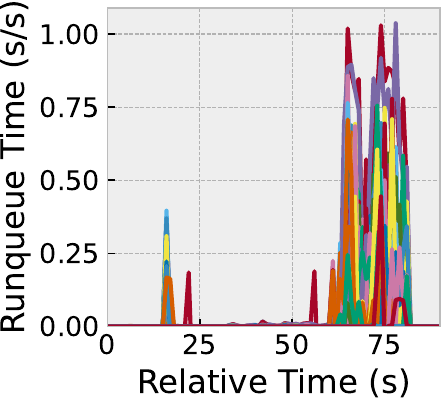}
    }\\ \upp\upp
    \subfloat[\label{fig:solr:epoll_wait}]{
      \includegraphics[width=0.31\columnwidth]{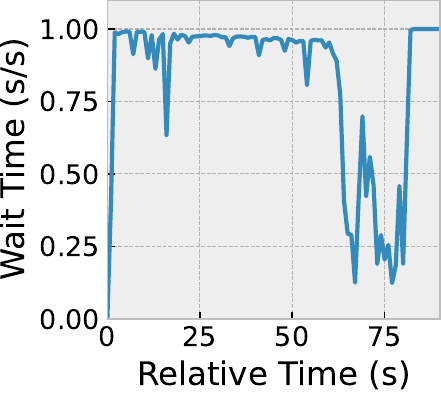}
    }
    \subfloat[\label{fig:solr:external_wait}]{
      \includegraphics[width=0.31\columnwidth]{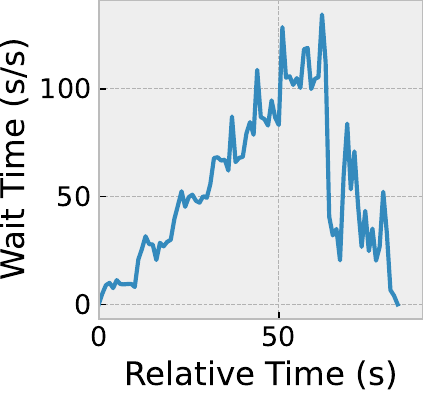}
    } \upp
    \caption{\textbf{Solr target metric observation and performance degradation deconstruction:} (a) User load pattern; (b) Solr $95^{th}$ percentile response time; (c) Solr threads runqueue activity; (d) Epoll wait time; (e) Combined connection wait time.} \upp\upp\upp
\end{figure}
\paragraph{\textbf{Solr}}\textbf{Experiment Scenario \& Target Metric Observation.} For \emph{Solr}'s experiment, we modified \emph{Cloudsuite's Solr} benchmark to allow variable load patterns. Figure \ref{fig:solr:users} shows the applied load pattern, while Figure \ref{fig:solr:target} displays the server's $95^{th}$ percentile \emph{response time}. After \emph{cache warm-up}, the server stabilizes at \SIrange[range-phrase=\,--\,]{2}{4}{\milli\second} response time. However, \SI{63}{\second} into the experiment, with $130$ users, the server becomes overwhelmed, and response time rises to \SI{100}{\milli\second}.

\textbf{Deconstructing Performance Degradation.} To deconstruct performance degradation, we start by examining application's scheduling activity. Figure \ref{fig:solr:rq_time} shows server's \emph{rq\_time}, which negatively affected the target metric. This suggests the server's degraded performance results from CPU contention due to insufficient CPU allocation for the container.

Despite runqueue activity coinciding with server degradation, we explore deeper connections to response time increase. \emph{Solr} uses a single epoll resource for connection multiplexing, shared by multiple waiting threads. Figure \ref{fig:solr:epoll_wait} shows these threads' total \emph{epoll\_wait\_time} for the same epoll \emph{BRI}, which dips significantly during performance degradation. Combining this with the application's connection's \emph{socket\_wait\_time} via \emph{epoll\_file\_wait} (Figure \ref{fig:solr:external_wait}), reveals \emph{rq\_time} activity impacting connection handling. All threads that waited for this epoll \emph{BRI} during degradation also showed anomalous runqueue activity in Figure \ref{fig:solr:rq_time}, suggesting the added latency from runqueue wait time impaired timely data reception and processing by connection threads.

\begin{figure}[t]
    \centering
    \upp\upp\upp
    \subfloat[\label{fig:ml:load}]{
      \includegraphics[width=0.31\columnwidth]{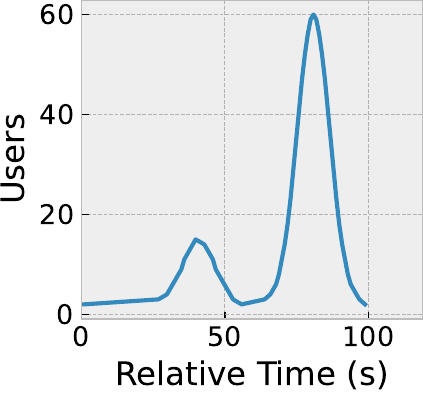}
    }
    \subfloat[\label{fig:ml:target}]{
      \includegraphics[width=0.31\columnwidth]{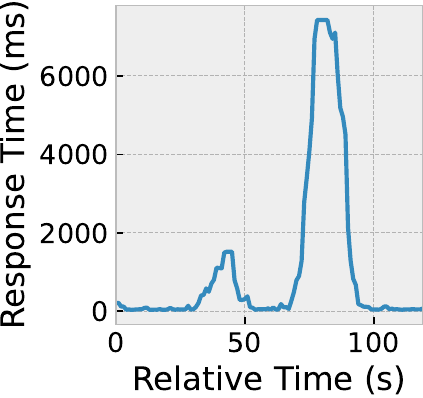}
    }
     \subfloat[\label{fig:ml:runtime}]{
      \includegraphics[width=0.31\columnwidth]{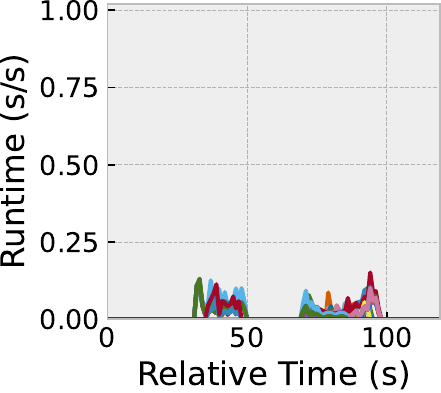}
    }\\ \upp\upp
    \subfloat[\label{fig:ml:rqtime}]{
    \upp\upp
      \includegraphics[width=0.31\columnwidth]{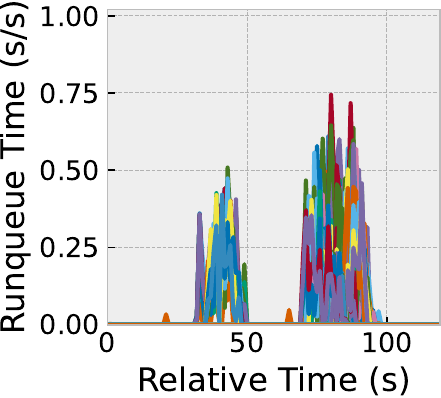}
    }
    \subfloat[\label{fig:ml:external_wait}]{
    \upp\upp
      \includegraphics[width=0.31\columnwidth]{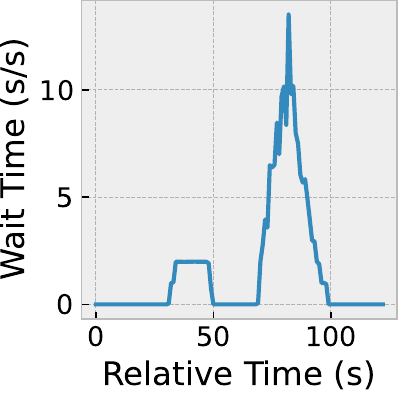}
    }
    \subfloat[\label{fig:ml:futex_wake}]{
    \upp\upp
      \includegraphics[width=0.31\columnwidth]{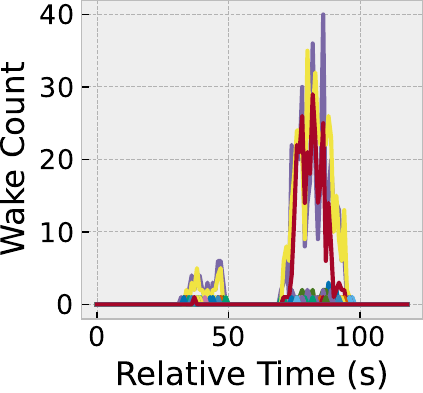}
    }\\ \upp\upp
    \subfloat[\label{fig:ml:futex_wait}]{
    \upp\upp
      \includegraphics[width=0.31\columnwidth]{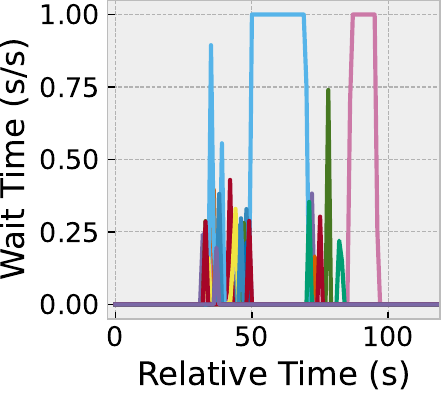}
    }
    \subfloat[\label{fig:ml:thread_sched}]{
    \upp\upp
      \includegraphics[width=0.31\columnwidth]{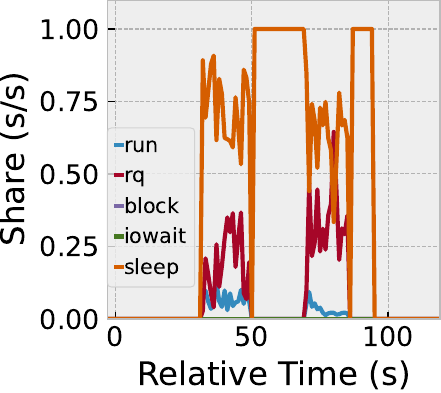}
    } \upp
    \cprotect\caption{\textbf{ML-inference target metric observation and performance degradation deconstruction:} (a) ML-inference load pattern; (b) ML-inference $95^{th}$ percentile response time; (c) ML-inference threads' \emph{runtime}; (d) ML-inference threads' \emph{rq\_time}; (e) ML-inference combined connection wait time; (f) Futex wake activity; (g) Thread \verb|101439| futex wait activity; (h) Thread \verb|101439| scheduling activity.} \upp\upp\upp
\end{figure}

\paragraph{\textbf{ML-inference}}\textbf{Experiment Scenario \& Target Metric Observation.} As described in Section~\ref{sec:experiments}, we tested ML-inference server deployed with \emph{gunicorn} ($4$ workers). \emph{Locust} generated a load pattern (Figure \ref{fig:ml:load}) between $2$-$60$ users to induce saturation. Figure \ref{fig:ml:target} shows the 95th percentile \emph{response time} latency, our target metric, with an increase factor of 100 (\SI{75}{\milli\second} to \SI{7500}{\milli\second}) while degraded.

\textbf{Deconstructing Performance Degradation.} To understand ML-inference performance degradation, we perform the following analysis of one of the four \emph{gunicorn} workers, which is generalisable to others. Figures \ref{fig:ml:runtime} and \ref{fig:ml:rqtime} show threads' \emph{runtime} and \emph{rq\_time} activities. Runtime (and runqueue) activity occurs only when load exceeds $2$ users, suggesting gunicorn load-balances connections among $4$ workers, with this worker idle when the workload is below $2$ users. Figure \ref{fig:ml:external_wait} further confirms this, showing the combined \textit{socket\_wait\_time} for client connections, aligning with worker's scheduling activity.

The worker uses epoll for multiple connections, with one thread as the entrypoint. This thread signals data arrival by waking threads waiting on futexes (Figure \ref{fig:ml:futex_wake}), suggesting their use as work schedulers. Figures \ref{fig:ml:futex_wait} and \ref{fig:ml:thread_sched} show a thread's \emph{futex\_wait\_time} and scheduling activity, respectively. The scheduling activity closely aligns with futex wakes triggered by the entrypoint thread upon new data arrival.

In summary, undesired runqueue activity indicates container CPU saturation caused the ML-inference server's degraded performance. However, this analysis further reveals the application's work distribution among its threads through its use of epoll and futex resources, which is crucial for identifying critical resources and potential architectural constraints.

\begin{figure}[t]
    \centering
    \subfloat[\label{fig:redis:target}]{
      \includegraphics[width=0.34\columnwidth]{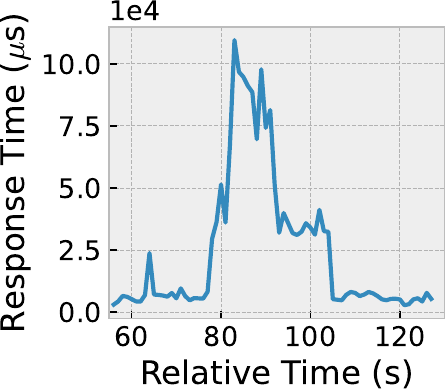}
    }
    \subfloat[\label{fig:redis:sched}]{
      \includegraphics[width=0.31\columnwidth]{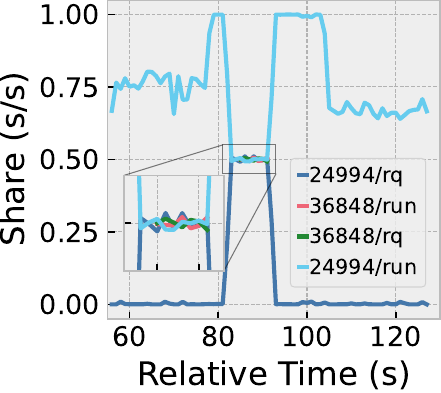}
    }
    \vspace*{-0.2cm}
    \caption{\textbf{Redis target metric observation and performance degradation deconstruction:} (a) Memtier benchmark $95^{th}$ percentile response time; (b) Redis \emph{runtime} and \emph{rq\_time} activity.} \upp\upp\upp
\end{figure}

\paragraph{\textbf{Redis}}\textbf{Experiment Scenario \& Target Metric Observation.} Redis experiment runs the \emph{memtier} benchmark~\cite{redis_memtier}, measuring its $95^{th}$ percentile latency (Figure \ref{fig:redis:target}). We introduced the official Redis benchmark~\cite{redis_benchmark} as an intervention, starting at \SI{76}{\second} and lasting \SI{30}{\second}, to saturate the server.

\textbf{Deconstructing Performance Deconstruction.} Analysing server scheduling (Figure \ref{fig:redis:sched}) reveals two key points. Initially, the \emph{Redis} main thread's runtime increases to $100$\% at the start of the benchmark. Soon after, another Redis thread contends for the same CPU resources. This contention lasts \SI{13}{\second} until the new thread terminates. Despite the main thread regaining uninterrupted execution, server remains overwhelmed. Post-intervention, thread core usage drops to $66$\%, adequately handling the workload. Scheduling activity, combined with main thread managing all external requests via epoll, suggests CPU core contention and saturation caused server's degradation.

\subsection{Externally Constrained}
\begin{figure}[t]
    \centering
    \upp\upp
    \subfloat[\label{fig:teastore:load}]{
      \includegraphics[width=0.31\columnwidth]{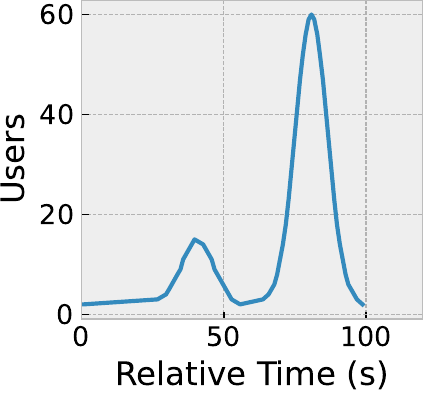}
    }
    \subfloat[\label{fig:teastore:target}]{
      \includegraphics[width=0.31\columnwidth]{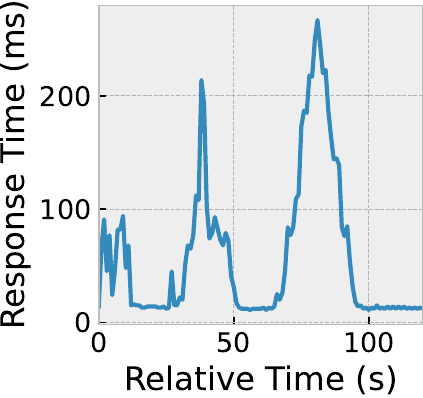}
    }
    \subfloat[\label{fig:teastore:rq_time}]{
      \includegraphics[width=0.31\columnwidth]{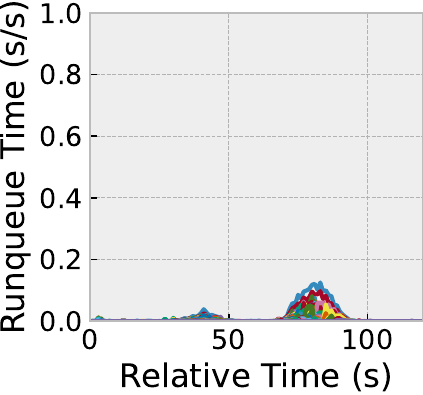}
    }\\ \upp\upp
    \subfloat[\label{fig:teastore:runtime}]{
    \upp\upp
      \includegraphics[width=0.31\columnwidth]{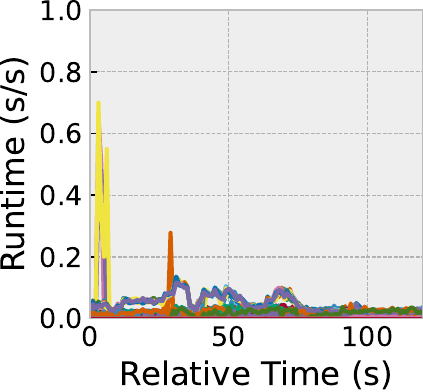}
    }
    \subfloat[\label{fig:teastore:histogram}]{
    \upp\upp
      \includegraphics[width=0.62\columnwidth]{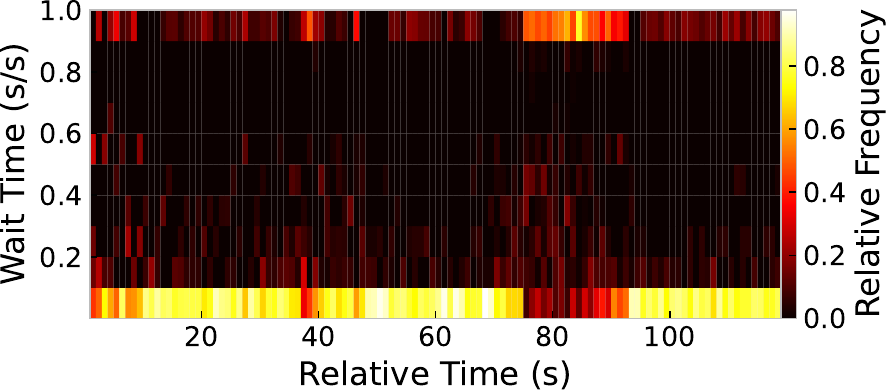}
    } \upp
    \caption{\textbf{Teastore target metric observation and performance degradation deconstruction:} (a) Teastore load pattern; (b) End-to-end $95^{th}$ percentile response time; (c) Webui's threads' \emph{rq\_time}; (d) Webui's threads' \emph{runtime}; (e) Timeseries histogram of webui's wait time for the constrained service.} \upp\upp\upp
\end{figure}

\paragraph{\textbf{Teastore}}\textbf{Experiment Scenario \& Target Metric Observation.} In previous sections, we primarily focused on bottlenecks within the analyzed application. In multi-service systems, however, external services can cause performance degradation. Our final experiment deploys \emph{Teastore} microservice system~\cite{von2018teastore}, monitoring \emph{webui} component while constraining the \emph{persistence} service. We apply a variable load pattern (Figure \ref{fig:teastore:load}) and measure the client's $95^{th}$ percentile response time (Figure \ref{fig:teastore:target}).

\textbf{Deconstructing Performance Degradation.} We analyze the system's \emph{rq\_time} and \emph{runtime} in Figures \ref{fig:teastore:rq_time} and \ref{fig:teastore:runtime}. The \emph{webui} component shows low CPU usage throughout, with runqueue activity increasing at peak load. This runqueue latency increase doesn't fully explain the sharp response time rise. Figure \ref{fig:teastore:histogram} presents a time series histogram of \emph{webui} service \emph{socket\_wait\_time} derived from connections to the constrained \emph{persistence} service. Higher y-axis positions indicate longer wait times, while bin brightness shows relative frequency within each timestamp column.

Figure \ref{fig:teastore:histogram} highlights three key regions. The most notable, \SIrange[range-phrase=\,--\,]{70}{92}{\second} interval, shows \emph{persistence} service wait times shifting from \SIrange[range-phrase=\,--\,]{0}{0.1}{\second} to \SIrange[range-phrase=\,--\,]{0.9}{1}{\second}, coinciding with peak load (i.\,e., $60$ users). Within the \SIrange[range-phrase=\,--\,]{35}{48}{\second} interval, Figure \ref{fig:teastore:histogram} also shows an increased wait time for \emph{persistence} service connections, also coinciding with an increase in user count ($15$ users). Lastly, the \SIrange[range-phrase=\,--\,]{0}{10}{\second}
interval, aligns with the initial high latency, despite a low user count (2 users). These insights indicate the degradation is external to the \emph{webui} service, demonstrating the selected metrics' utility in identifying both internal and external sources of contention.

\section{Discussion}\label{sec:discussion}
\paragraph{\textbf{Resource Contention Identification}} Metrics presented in Table \ref{table:metrics} enable pinpointing potential sources of interference, after identifying the resources constraining an application, as demonstrated in Section~\ref{sec:results}. For instance, when disk contention is the root cause, metrics collected at a \textbf{"global"} scope (see Section~\ref{sec:archi}) reveals the interference caused by co-located processes. However, some resources require additional observability. Currently, memory contention manifests itself as increased \emph{runtime} due to more cycles per instruction (as observed in \emph{Cassandra}'s evaluation, Figure~\ref{fig:cassandra:second_runtime}), where \emph{runtime} serves as a proxy for the real contention source. Hence, distinguishing load increase from memory bandwidth requires instrumenting additional CPU performance counter metrics, which we plan to address in our future work.

\begin{table}
\upp
\centering
\caption{Instrumentation overhead: Latency mean and stdev with(out) instrumentation.}
\upp
\label{table:overhead}
\setlength{\tabcolsep}{0.5pt}
\resizebox{\columnwidth}{!}
{
\begin{tabular}{|c|c|c|c|}
\hline
\textbf{App} & 
\multicolumn{1}{c|}{\textbf{\begin{tabular}[c]{@{}c@{}}Latency\textsubscript{w/out}\\ Mean ± SD (ms)\end{tabular}}} & 
\multicolumn{1}{c|}{\textbf{\begin{tabular}[c]{@{}c@{}}Latency\textsubscript{with}\\ Mean ± SD (ms)\end{tabular}}} & 
\multicolumn{1}{c|}{\textbf{\begin{tabular}[c]{@{}c@{}}Latency \\ Overhead (ms)\end{tabular}}} \\
\hline
Redis       &  $2.053 \pm 0.155$ & $2.425 \pm 0.310$ & $0.372$ \\ \hline
Cassandra   &  $1.689 \pm 0.002$ & $1.758 \pm 0.007$ & $0.070$ \\ \hline
MySQL       & $47.193 \pm 2.344$ & $47.495 \pm 3.023$ & $0.302$ \\ \hline
Kafka       &  $2.170 \pm 0.051$ & $2.231 \pm 0.033$ & $0.061$ \\ \hline
\end{tabular} 
} 
\upp\upp\upp\upp
\end{table}

\paragraph{\textbf{Instrumentation Overhead}}
We evaluate overhead for four standalone applications using the same workloads from Section~\ref{sec:experiments}, comparing latency with(out) our instrumentation over $10$ runs. Table~\ref{table:overhead} shows the latency mean and standard deviation for both scenarios. Induced instrumentation overhead depends on the frequency and computation of our \emph{eBPF} programs and varies with each application's kernel resource needs. We observe that instrumentation impacts Cassandra and Kafka least ($\approx$ \SI{0.07}{\milli\second}) and Redis and MySQL most ($\approx$ \SI{0.3}{\milli\second}), likely because MySQL and Redis process requests immediately, while Kafka and Cassandra defer processing.

\section{Conclusion}\label{sec:conclusion} 
This paper introduced and implemented $16$ \emph{eBPF-based} instrumentation metrics to support application-agnostic performance degradation diagnosis. Unlike existing diagnosis approaches that overlook statistics on thread interactions with specific kernel resources (such as CPU, disks, and locks), our instrumentation approach collects metrics at this level of granularity, enabling detailed QoS degradation diagnosis. 
Results show that for a representative group of online data-intensive applications under stress from variable workload patterns and resource contention scenarios, the implemented \emph{eBPF-based} metrics enable effectively deconstructing the causes of their performance degradation. As our future work, we plan to extend the instrumented metrics to further diagnose sources of memory contention induced by co-located processes.
\bibliographystyle{IEEEtran}
\bibliography{bibliography}

\end{document}